\PassOptionsToPackage{utf8}{inputenc}
\documentclass{bioinfo}
\copyrightyear{2015} \pubyear{2015}

\access{Advance Access Publication Date: Day Month Year}
\appnotes{Manuscript Category}

\usepackage[colorlinks,
            linkcolor=blue,
            anchorcolor=blue,
            citecolor=blue]{hyperref}
\usepackage{booktabs}
\usepackage{multirow}
\usepackage{graphicx}
\usepackage{threeparttable}

\begin{document}
\firstpage{1}

\subtitle{Subject Section}

\title[short Title]{AMGC:Adaptive matching based genome compression}
\author[Sample \textit{et~al}.]{Jia Wang\,$^{\text{\sfb 1}}$, Yi Niu\,$^{\text{\sfb 1,2}*}$, Tianyi Xu\,$^{\text{\sfb 1}}$, Mingming Ma\,$^{\text{\sfb 1}}$, Dahua Gao\,$^{\text{\sfb 1}}$  and Guangming Shi\,$^{\text{\sfb 1,2}}$}
\address{$^{\text{\sf 1}}$School of artificial intelligence, Xidian University, Xian, China 710071 and \\
$^{\text{\sf 2}}$ The Pengcheng Lab, Shenzhen, China,518055.}

\corresp{$^\ast$To whom correspondence should be addressed.}

\history{xxx}

\editor{xxx}

\abstract{\textbf{Motivation:} Despite significant advances in Third-Generation Sequencing (TGS) technologies, Next-Generation Sequencing (NGS) technologies remain dominant in the current sequencing market. This is due to the lower error rates and richer analytical software of NGS than that of TGS. NGS technologies generate vast amounts of genomic data including short reads, quality values and read identifiers. As a result, efficient compression of such data has become a pressing need, leading to extensive research efforts focused on designing FASTQ compressors. Previous researches show that lossless compression of quality values seems to reach its limits. But there remain lots of room for the compression of the reads part. \\
\textbf{Results:} By investigating the characters of the sequencing process, we present a new algorithm for compressing reads in FASTQ files, which can be integrated into various genomic compression tools. We first reviewed the pipeline of reference-based algorithms and identified three key components that heavily impact storage results: the matching positions of reads on the reference sequence(\textit{refpos}), the mismatched positions of bases on reads(\textit{mispos}) and the matching failed reads(\textit{unmapseq}). To reduce their sizes, we conducted a detailed analysis of the distribution of matching positions and sequencing errors and then developed the three modules of AMGC. According to the experiment results, AMGC outperformed the current state-of-the-art methods, achieving an \textbf{81.23\%} gain in compression ratio on average compared with the second-best-performing compressor. \\
\textbf{Availability:} \href{https://github.com/wj-inf/AMGC}{https://github.com/wj-inf/AMGC}\\
\textbf{Contact:} \href{niuyi@mail.xidian.edu.cn}{niuyi@mail.xidian.edu.cn}\\}
\maketitle

\section{Introduction}
Since the appearance of Next-Generation Sequencing (NGS) technologies (\cite{mardis2008next}), we have witnessed the rapid development of sequencing technologies. Although Third-Generation Sequencing (TGS) technologies (\cite{schadt2010window}) have been widely developed, NGS technologies remain dominant in the current sequencing market for two main reasons. Firstly, the sequencing data produced by NGS technologies has a significantly lower error rate than that of TGS technologies. For example, in the HiSeq (NGS technology) data, 92\% of sequenced bases had a base quality score $\geqslant$ 30, which means that the estimated error rate was less than 0.1\% (\cite{ma2019analysis}). TGS long reads usually have high error rates: ~15\% with PacBio sequencing, and as high as ~40\% with Oxford Nanopore sequencing (\cite{wee2019bioinformatics}). These high error rates make the assembly of TGS sequences seem disproportionally complex and expensive compared to the assembly of NGS sequences. NGS data also has richer analytical software. Tools like SAM tools (\cite{li2009sequence}), BWA-MEM and Bowtie2 (\cite{li2013aligning}) have been widely used, making downstream analysis of NGS data more convenient. As a result, well-known companies Illumina and BGI still regard NGS as an important sequencing business, especially Illumina's Hiseq sequencer series.

NGS technologies can generate massive amounts of sequencing data every day. For example, a \href{https://www.illumina.com.cn/systems/sequencing- platforms/novaseq/specifications.html}{NovaSeq 6000} sequencer can produce up to 3TB data in a single 44-hour run. However, the expansion of NGS data poses a serious challenge to its storage and transfer. How to effectively store and transmit these massive high-throughput genomic data has been a pressing issue. Genomic data compression technology becomes an important way to solve this problem. As there exist clear formats for read identifiers, the compression of NGS data focuses on short reads and quality values. According to previous research (\cite{niu2022aco}), lossless compression of quality values seems to reach its limits. However, the compression of the reads part exists further potential.

In this work, we propose an algorithm for read compression, AMGC, which improves over the state-of-the-art compressors. AMGC consists of three key modules, aiming at solving three encoding redundancies in the current pipeline of reference-based approaches. The first module focus on the encoding of reference matching positions (\textit{refpos}). By further analyzing the distribution of \textit{refpos}, we discovered that adjacent reads are more likely to match close positions on the reference sequence and propose the bit-plane-based differential \textit{refpos} encoding module. After that, we investigate the synthesis reaction involved in the NGS sequencing process and find that sequencing errors are more common in the trailing part of the short reads. This results in a uniform distribution of sequencing errors. To address this, we propose the adaptive binary mismatching positions encoding module. It uses a binary vector to denote matching mistakes and then encodes the vector using a high-order context. Finally, as the front subsegment has better quality than the rear subsegment, continuous mismatch bases in the trail may lead to the match fault of the whole read. Accordingly, we further propose the recursive split matching module. Read will be split into two segments after matching fault, and the two segments are aligned again. So the high-quality front subsegment could match successfully.

We evaluated AMGC on the datasets chosen from the standard test datasets provided by \href{http://www.avs.org.cn/en/}{AVS-G}. The experiment results showed that AMGC got an 81.23\% average gain compared with the second-best-performing compressors with comparable RAM and time usage. 

\section{Related works}
NGS technologies have produced large numbers of short genomic reads that are highly redundant and compressible. However, popular general-purpose compression tools such as gzip, bzip2, and 7z can not achieve satisfactory performance on genome data. Because they do not take full advantage of the biological properties of the data, like repetitive fragments and palindromes. Therefore, efficient compression methods designed specifically for genome data are highly needed, bringing lots of significant work.

For the original FASTQ format, several specialized compression methods have been proposed (\cite{bonfield2013compression}) (\cite{deorowicz2011compression}). Based on whether additional reference genomes are required, these compression techniques can be broadly categorized into three groups: High-order context compression algorithms, assembly-based compression algorithms, and reference-based compression algorithms. High-order context compression algorithms are progressively being phased out. 

Assembly-based approaches exploit the similarity between reads in FASTQ files and splice the measured short sequences into a long sequence by the assembly algorithm. They can deal with unsequenced species and macrogenomics, as they do not require a reference genome. Quip (\cite{jones2012compression}) uses a Bloom filter to construct a de Bruijn graph for splicing. And the spliced contigs sequences are used as the reference genome for sequenced reads in the original FASTQ file. Finally, only the aligning position and variation information of each sequenced read are recorded for coding. Leon (\cite{benoit2015reference}) constructs a de Bruijn graph using all sequenced reads. Similar to the Quip method, Leon also uses a Bloom filter to filter out the low-abundance k-mer nodes to reduce the number of bifurcation nodes. Each sequenced read is represented as the path start node plus the fork node information. Then it is encoded in a tree structure and a zero-order context model, respectively. HARC (\cite{chandak2018compression}) reorders reads roughly according to their position in the genome. These reordered reads are encoded to remove the redundancy between consecutive reads, and the parameters obtained are stored in different files. The files obtained above are compressed using Lempel-Ziv (\cite{ziv1977universal}) and BWT (\cite{burrows1994block}) based universal compressors. PgRC (\cite{kowalski2020pgrc}) constructs pseudogenomes, which are approximately the shortest common superstrings of reads. Then it encodes the reads according to their mapped positions on the pseudogenomes.

Reference-based approaches align the sequenced reads to an external reference sequence to identify similarities between them. Only the position of the comparison and variation information is encoded instead of the original target sequences. These methods achieve better performance by exploiting the similarity between the sequenced genome and the reference genome. Using the human genome as an example, over 99\% of base pairs can be matched to the reference genome (\cite{Lander}). In LW-FQZip (\cite{zhang2015light}), reads are preprocessed by a lightweight mapping model and then mapping results are compressed by a general-purpose tool such as LZMA. Algorithms such as CRAM (\cite{fritz2011efficient}) employ a standard reference-based scheme to compress SAM/BAM data. This scheme relies on a highly optimized SAM field that supports various statistical models. The use of these models results in a reduction of the compression rate, as they can more efficiently capture the relevant information from the reference sequence. GTZ (\cite{xing2017gtz}) is a widely preferred reference-based algorithm in industrial applications. It performs significantly better than other previously proposed reference-based algorithms in terms of compression results.

However, all the current algorithms suffer from three drawbacks. A widely used assumption is that \textit{refpos} distribute uniformly across the reference sequence, which results in the use of fixed bit coding for \textit{refpos} compression. Another widely used assumption is that the distribution of mismatching positions (\textit{mispos}) is also uniform. So they use difference to encode \textit{mispos}. Finally, they do not consider the influence of continuous mismatching and design a simple matching model. The model sets a threshold for the mismatching count of the entire read. However, further analysis of the data reveals that both the distribution of \textit{refpos} and \textit{mispos} were non-uniform. It means all three models used before are improvable. Based on these, AMGC designed three modules targeting improving the three disadvantages.


\section{Insight}
Before the detailed introduction of the three modules of the proposed AMGC algorithm, we reveal two important properties of NGS sequencing process. It will help the reader to understand the technical motivation of the three key modules. In Section 3.1, we first review the pipeline of reference-based algorithms taking the integer-mapped k-mer indexing method (available at http://www.ysunlab.org/kic.jsp) as an example. And then we show the effect of each matching result on the final compressed size. In Sections 3.2 and 3.3, we simply review the process of NGS sequencing and the reaction of the synthesis, followed by a detailed analysis of the distribution of \textit{refpos} and \textit{mispos}.

\subsection{Pipeline of reference-based approaches}

\begin{figure*}
\centering
\includegraphics[width=1.0\textwidth]{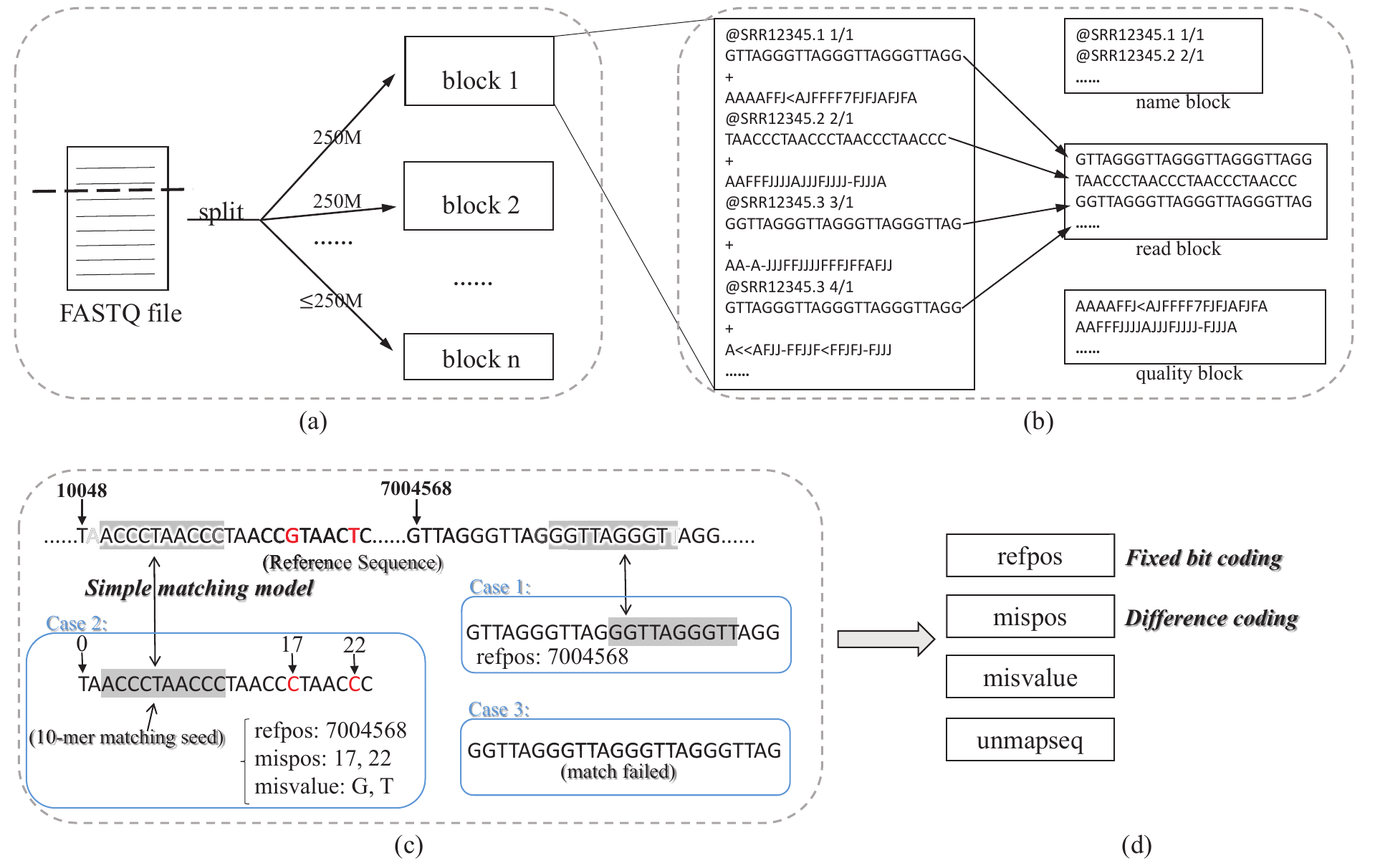}
\caption{The pipeline of reference-based algorithms. The bold italics are the three modules other compressors used, and AMGC’s three modules are designed for improving them. The three blue boxes in (c) show the three cases of matching.}
\label{fig1}
\end{figure*}

Notation: Let "\textit{refpos}" represent the matching positions of each read on the reference sequence. Let "\textit{mispos}" refer to the position of mismatching bases on read. Let "\textit{misvalue}" represent the correct base on the reference sequence for each mismatch position. Let "\textit{unmapseq}" refer to the set of reads that failed to match.

To begin, a FASTQ file is split into multiple 250MB memory blocks(as shown in Fig. \ref{fig1}(a)). This is necessary because FASTQ files are typically too large to be loaded into RAM at once. And splitting them enables easier multi-threaded processing. In Fig. \ref{fig1}(b), each 250MB block is further divided into "name block", "read block", and "quality block". Each 250MB block is unique, thus requiring adaptive encoding solutions for reads within different blocks.

Our focus is on the “read block”. In Fig. \ref{fig1}(c), each read is aligned to the reference sequence one by one using the 10-mer integer model. Some read match perfectly (Case 1), and we only need to record their \textit{refpos}. Some other reads match successfully but with a few different bases (Case 2), and we need to record their \textit{refpos}, \textit{mispos}, and \textit{misvalue}. For the remaining reads (Case 3), the part in \textit{unmapseq}, the match failed due to their large gap from all substrings in the reference sequence. More specifically, after determining the reference position by k-mer, it will expand to both ends and compare the bases at each corresponding position. If the number of different bases exceeds a certain threshold, the signal time alignment will be judged as a failure. When multiple k-mer mappings fail, read is judged as a match failure. After the matching process, the read block is transferred to the four parts in Fig. \ref{fig1}(d), and the only thing left is compression.

\begin{figure}
\centering
\includegraphics[width=0.5\textwidth]{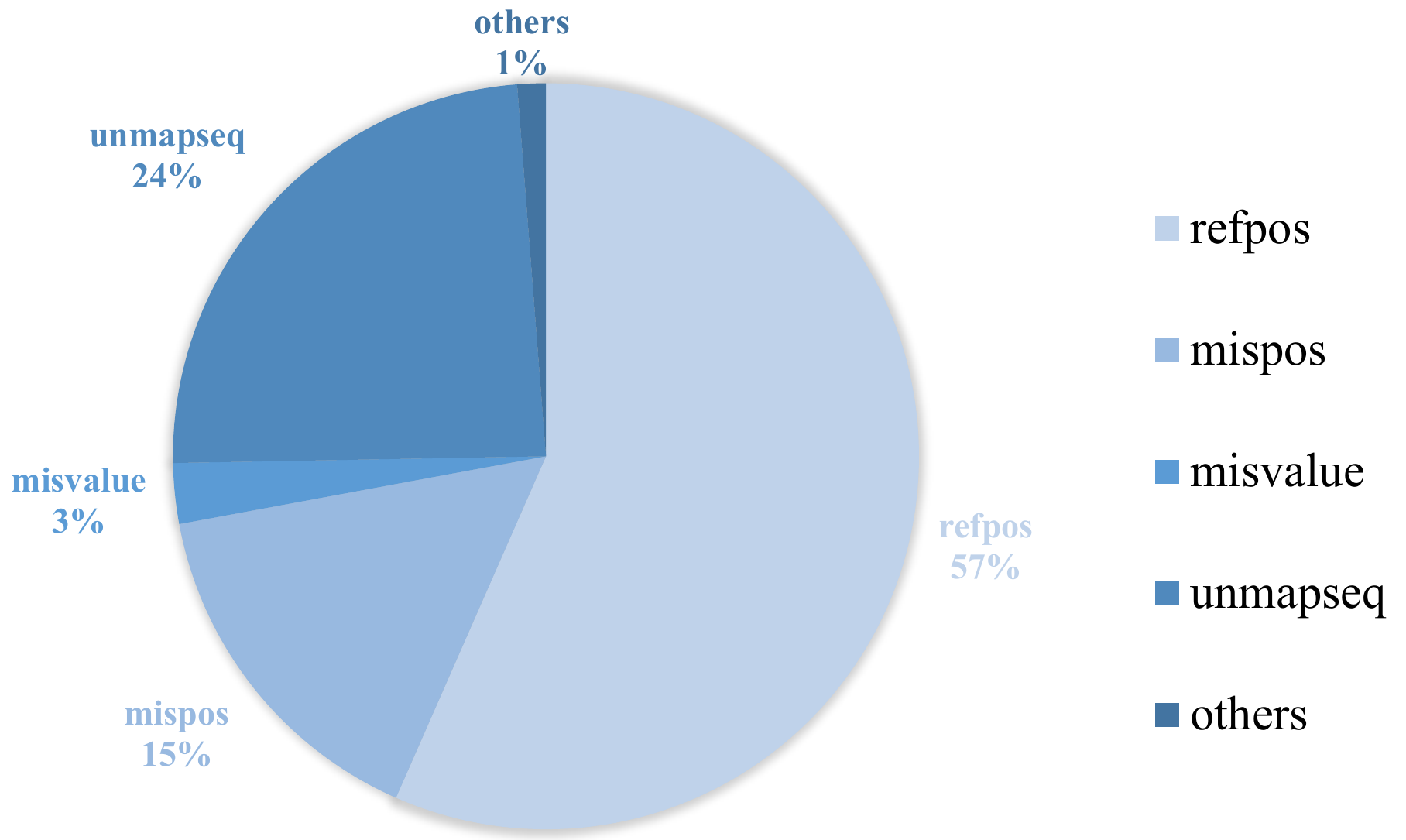}
\caption{The percentage of the size of each part in a read block after matching. The "others" category is about the processing of merger bases. The data used is from the first 50MB of data in the SRR6691666\_1.fastq processed by AVS-G.} 
\label{fig2}
\end{figure}

Fig. \ref{fig2} presents the proportion of each part in a read block after matching. As shown, the "\textit{refpos}," "\textit{unmapseq}," and "\textit{mispos}" components are critical factors in determining the size of the compressed data. To address this, AMGC has developed three modules to decrease compression size. The bit-plane-based differential \textit{refpos} encoding module significantly reduces the size of "\textit{refpos}" part. The recursive split matching module increases the number of successfully matched reads, thereby indirectly reducing the "\textit{unmapseq}" part. Additionally, the adaptive binary \textit{mispos} encoding module provides a new encoding method for "\textit{mispos}".

\subsection{Distribution of \textit{refpos}}

In the case of NGS, for example, the extracted DNA sample needs to be cut into small fragments during the sample preparation stage, then PCR amplifies the sample and aggregates it into Cluster before starting the synthetic sequencing session. The other strand paired with a fluorescently labelled deoxyribonucleotide of a different colour emits light signals of a different colour at the same time. These light signals are captured by the sequencer, which distinguishes different bases based on the different light signals and writes the sequencing results into a text file, such as a FASTQ file, to complete the sequencing process. The point is that as nearby small DNA fragments are sequenced, reads with similar positions in the FASTQ file are often matched to regions with similar positions, resulting in a strong local correlation in the comparison position information. Furthermore, since sequencing is done sequentially, the \textit{refpos} of the reads in the FASTQ file are growing locally linearly too.

\begin{figure}
\centering
\includegraphics[width=0.5\textwidth]{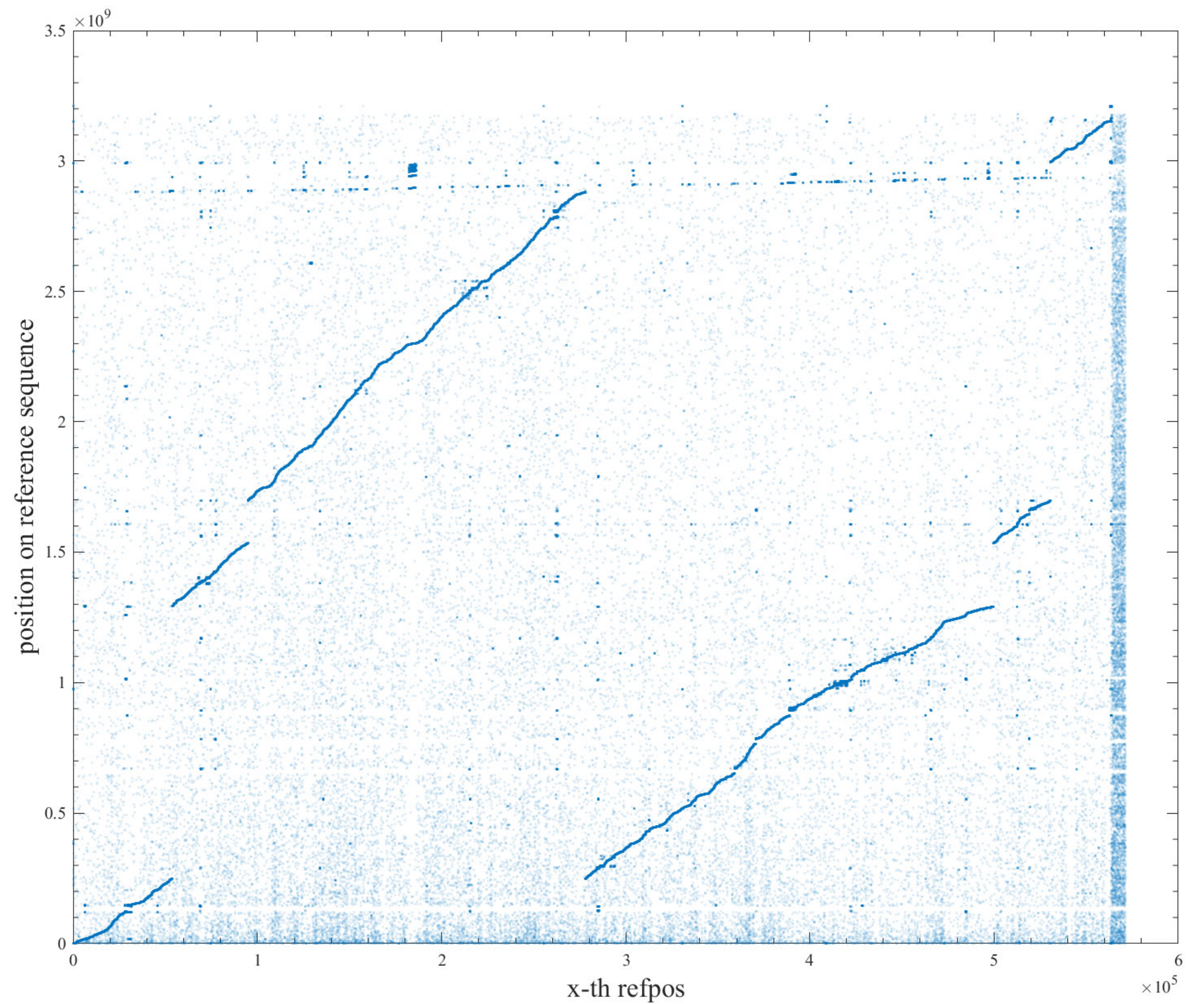}
\caption{Distribution of \textit{refpos} for all the reads of FASTQ file in a Scatter plot. The horizontal axis represents each read, while the vertical axis represents the position in the reference file hg38.fa.} 
\label{fig3}
\end{figure}

As shown in Fig. \ref{fig3}, a scatter plot of \textit{refpos} for all the reads of SRR6178157 is presented. It can be observed that \textit{refpos} is a one-dimensional signal with a locally growing trend and some noise. Hence, we can leverage this growth property to develop our \textit{refpos} encoding module.

\begin{figure}
\centering
\includegraphics[width=0.5\textwidth]{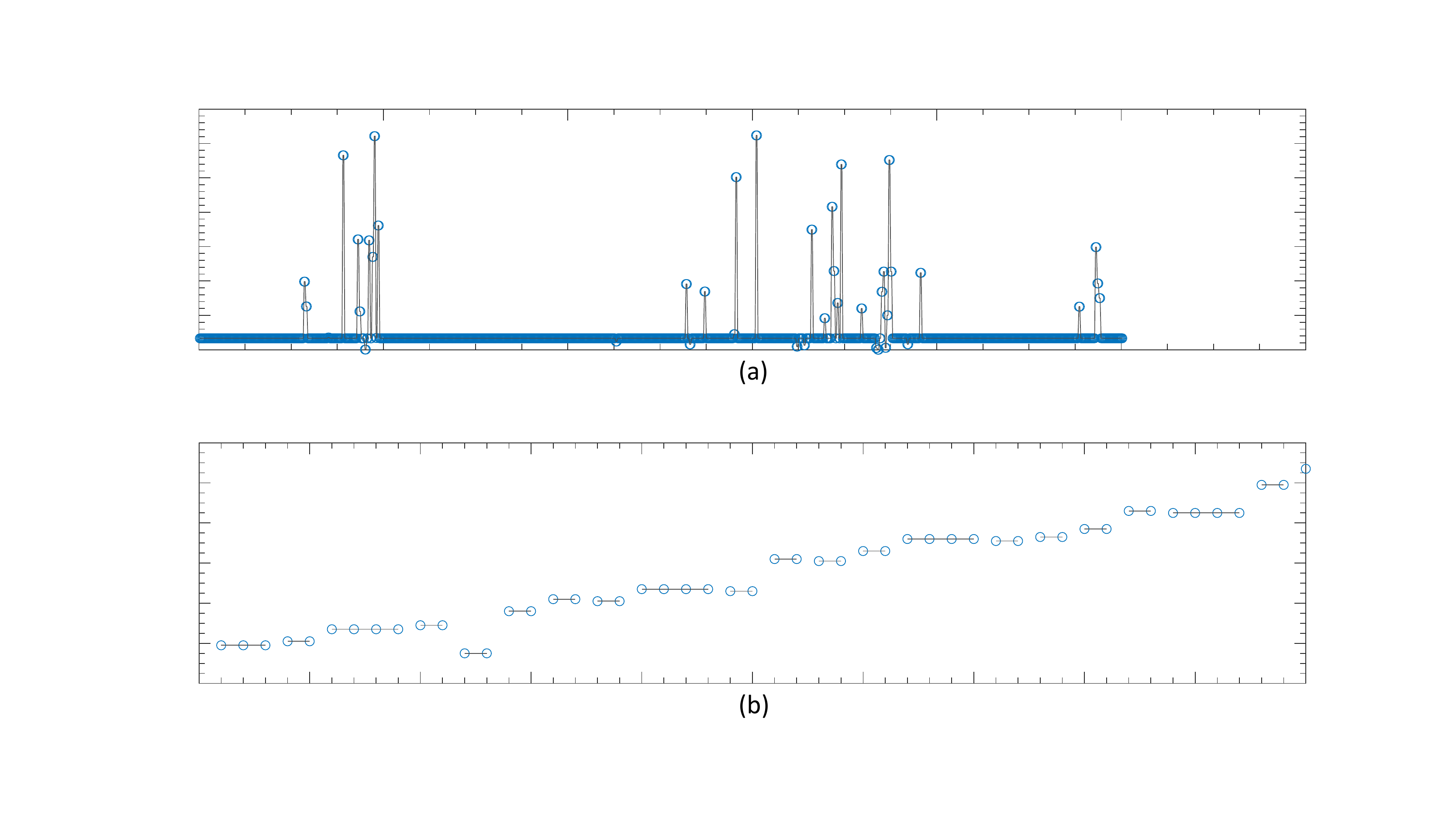}
\caption{Distribution of \textit{refpos} for local reads of FASTQ file in a Scatter plot} 
\label{fig4}
\end{figure}

To have a closer view, we zoomed in on Fig. \ref{fig3}. As depicted in Fig. \ref{fig4}, each blue scatter represents a \textit{refpos}, and the black line is used to aid observation. With 1000 consecutive refpos in Fig. \ref{fig4}(a), it can be observed that the signal grows linearly while displaying some noise at a distance. With 50 consecutive \textit{refpos} in Fig. \ref{fig4}(b), the black line segments indicate the same \textit{refpos} in succession. The occurrence of consecutive identical \textit{refpos} may result from the PCR process. These distribution characteristics of the \textit{refpos} signal serve as the basis for AMGC's design of the bit-plane-based differential \textit{refpos} encoding module.

\subsection{Distribution of \textit{mispos}}

Sequencing errors can introduce inaccuracies when comparing the resulting reads to the reference bases. To design a more effective compression algorithm, it's important to analyze the sources of errors. During the early stages of the synthesis and sequencing process, the reaction may be unstable, but the quality of the DNA enzyme is good. However, as sequencing progresses, the reaction stabilizes while the enzyme activity and specificity gradually decline, leading to cumulative errors that increase over time. This trend results in a slight decrease in the error rate of base sequencing followed by an increase after stabilization. Further analysis of this phenomenon is available in (\cite{minoche2011evaluation}).

\begin{figure}[h]
\centering
\includegraphics[width=0.5\textwidth]{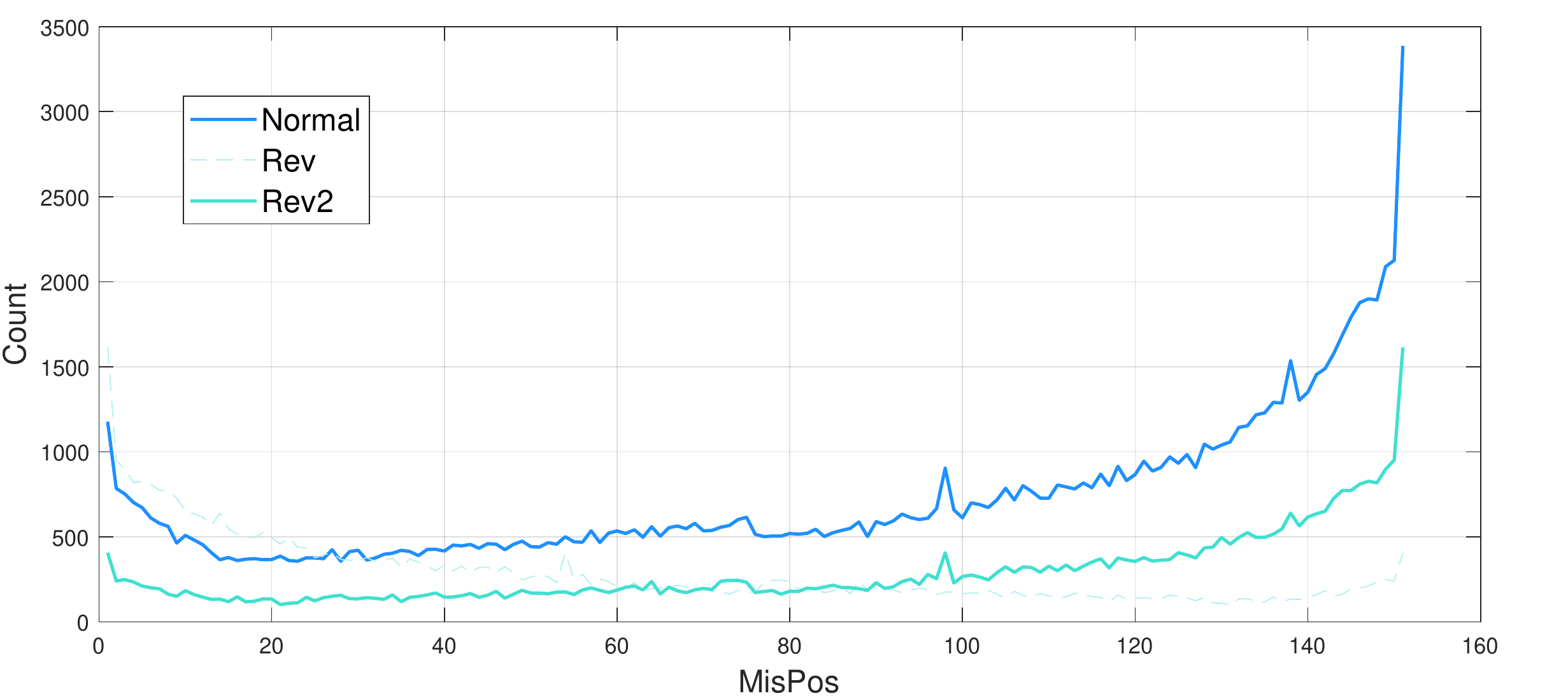}
\caption{Distribution of \textit{mispos} for the read in a Scatter plot. The horizontal axis indicates each position, while the vertical axis displays the number of errors. The two different lines, 'Normal' and 'Rev', represent the sequential and reverse comparisons, respectively. We also display 'Rev2', which is the reverse of 'Rev'.} 
\label{fig5}
\end{figure}

Fig. \ref{fig5} displays the distribution of \textit{mispos} in a scatter plot for 143003 reads in the FASTQ file “SRR6178157.fastq”. It clearly shows the trend analyzed above. To address this feature, we design the adaptive binary \textit{mispos} encoding module. We also carefully consider the effects of these mismatched bases on the matching process and come out with the recursive split matching module.

\section{Methods}

After providing the technical motivation for the three modules, this section delves into their implementation details. In Section 4.1, we describe the entire process of \textit{refpos} processing, including the use of deduplication, differential, and median filtering to fit the observed data features of \textit{refpos}. In Section 4.2, we introduce the adaptive binary \textit{mispos} encoding module, which uses a binary vector to represent matching mistakes and then uses a high-order context to encode the vector. In Section 4.3, we introduce the recursive split matching module to enable high-quality subsegment on unmapped reads matches.

\subsection{Bit-plane-based differential \textit{refpos} encoding}

Existing approaches do not consider the distribution of \textit{refpos}. We use the difference operation to account for the significant local correlation of the \textit{refpos}. Median filtering is used in the differencing process taking into account the noise points. The median filtering difference formula is :
$$
d\left( x \right) =p\left( x \right) \,\,-\,\,mid\left\{ p\left( x-1 \right) ,p\left( x-2 \right) ,p\left( x-3 \right) \right\} 
$$

where $p(x)$ represents the x-th \textit{refpos} and $d(x)$ represents the x-th \textit{refpos} after differencing, “$mid{}$” represent the median of three numbers, $p(0)$, $p(-1)$ and $p(-2)$ are zeros. We will first eliminate consecutive repetitions of the \textit{refpos} to fit the character that the same \textit{refpos} will display continually. 

The input for this technology is the “\textit{refpos}” part after matching. As an example, we use the human genome and hg38.fa, where each \textit{refpos} represents 32 bits. Here are the steps involved in this technology:

\begin{itemize}
    \item{Step 1: Remove consecutive duplicates of \textit{refpos}. Our analysis revealed that there were numerous instances of some consecutive \textit{refpos} being identical. This might be due to two identical readings being combined, and the duplicates were eliminated in the initial stage of processing the \textit{refpos}. We put the description of the duplicates in the “PosEqual” file, where 1 denotes that it is the same as the prior \textit{refpos}, and 0 denotes that it is different. To encode “PosEqual”, we use a binary arithmetic encoder.}
    \item{Step 2: Perform median filtering difference on the processed \textit{refpos}. In Section 3, we observe the linear growth property of refpos due to the DNA chain order sequencing. To process this one-dimensional signal, we apply differencing. Due to the presence of noise, the difference is disturbed and we use the median of the first three \textit{refpos} of the current \textit{refpos} as the subtracted number to filter out the noise.}
    
\begin{figure}[h]
\centering
\includegraphics[width=0.5\textwidth]{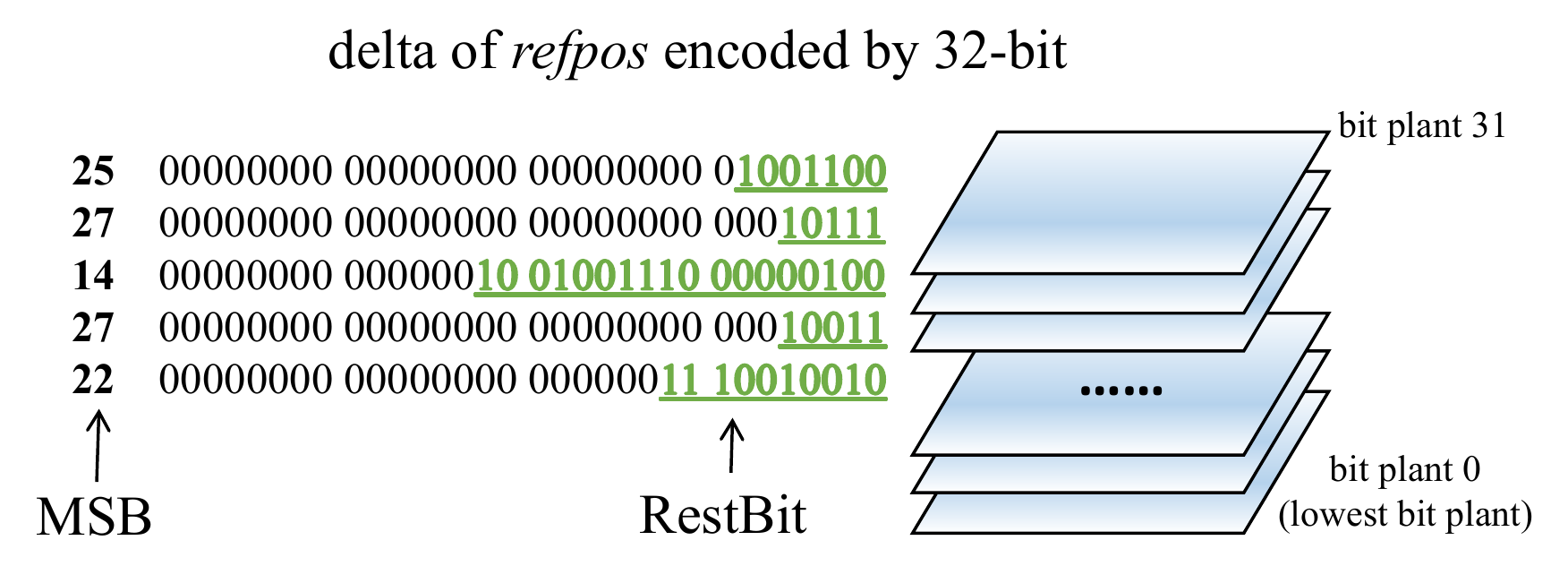}
\caption{Encoding algorithm of \textit{refpos} delta. The right figure shows what is bit plant. The left figure shows how to separate \textit{refpos} delta to "MSB" (most significant bit) and "RestBit"} 
\label{fig6}
\end{figure}

    \item{Step 3: Encode the differential result into bit planes. We represent the diff result as a positive or negative sign and 32-bit planes, which are split into “MSB” (most significant bit) and "RestBit” parts (Fig. \ref{fig6}). The sign part will be put into a simple arithmetic encoding model. For the “MSB” part, we use the 32-value arithmetic encoder to encode directly. For the “LeftBit” part, the MSB and the bit plane where each bit is located are used as the context when encoding.}
\end{itemize}

\subsection{Adaptive binary \textit{mispos} encoding}

Considering the distribution of \textit{mispos} on read is not uniform, we use a higher-order context encoding to capture the relevance of the mismatch trend. The input of this technology is the “\textit{mispos}” part after matching. We use 151-mer read as an example.

\begin{itemize}
    \item{Step 1: A binary vector is utilized to denote a matching mistake. According to the analysis in Section Insight, the error probability of each base in the read is not uniform, and the error probability of the tail part is higher. Therefore, we separate each position and then encode the error positions with the binary signal as shown in Fig. \ref{fig7}.}
    
\begin{figure}[h]
\centering
\includegraphics[width=0.5\textwidth]{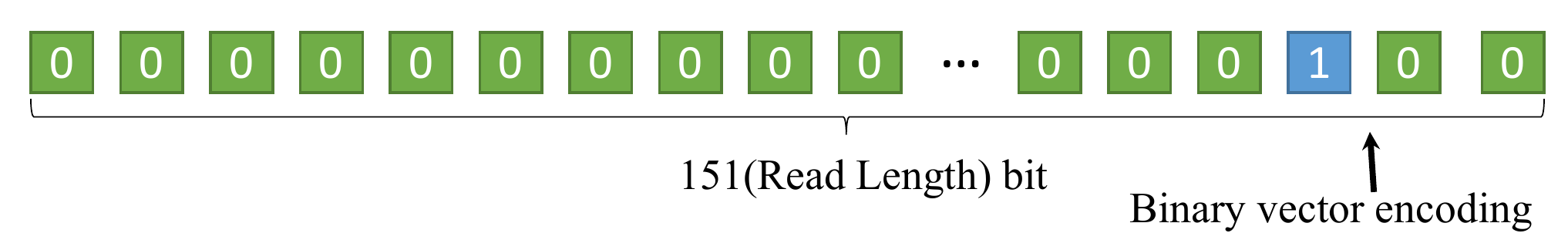}
\caption{Binary signal of \textit{mispos}} 
\label{fig7}
\end{figure}

    \item{Step 2: Use higher-order context encoding. To capture the relevance of the mismatch trend, instead of simply encoding each bit with the position as context, AMGC employs the number of errors in the first 10 bases of the current bit position as context. The probabilistic way of modelling the current encoded bit:
    $$p\left( x_i\left| \sum_{k=i-10}^{i-1}{x_k} \right. \right)$$}
\end{itemize}

\subsection{Recursive divide matching}
This technology is used in the matching process. Taking the integer-mapped k-mer indexing method as an example, mapping tools will match the read into the reference sequence through the short k-mer on read, and then expend k-mer to the entire read. If the number of mismatched bases exceeds the threshold, expending ends and matching fail. However, as analyzed in section 3.3, sequencing errors are more likely to occur at the end of the read. That means the high-quality segment in the front of the read is also thrown into the \textit{unmapseq} part, causing a waste of reference sequence. 

To address this, our recursive divide matching technology cuts the matching failed read into two equal segments and puts them into mapping tools again until they are too short to split. The splitting process runs recursively until the subsegment is shorter than a threshold.

\begin{figure}[h]
\centering
\includegraphics[width=0.47\textwidth]{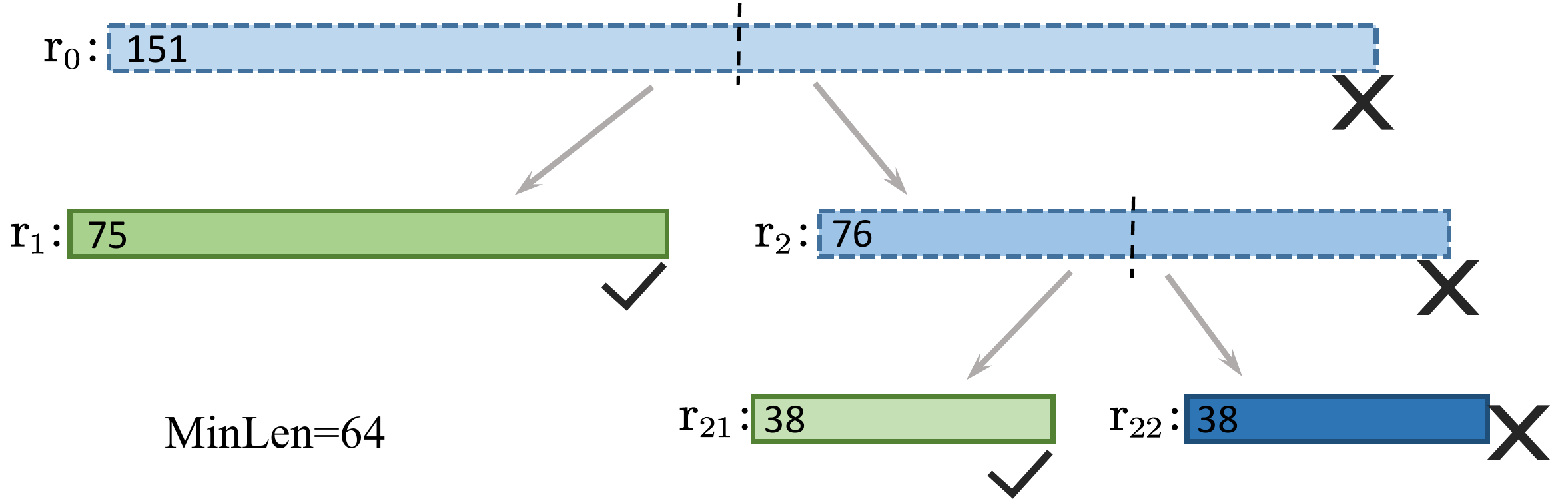}
\caption{Matching case of input read and its subsegments. Everyone will only be split into two equal-length parts from the middle. The nation, “right” or “wrong” represents matching success or failure. “MinLen” is the threshold stopping recursive dividing.} 
\label{fig8}
\end{figure}

As shown in Fig. \ref{fig8}, the input read (r0) has a length of 151. After its matching failed, we split it into two equal-length parts (r1 with length 75 and r2 with length 76). As the sequencing errors are more likely to occur at the end of the read, r1 matches successfully and r2 matches fail. We then divide r2 into r21 (length 38) and r22 (length 38). r21 matches successfully and r22 matches fail, but r22 is too short to divide further. The whole matching of r0 ends and we successfully decrease the unmapped base count and reduce the size of \textit{unmapseq} by making better usage of the reference sequence.

\section{Results and Discussion}

\begin{table*}[t]
\caption{ Datasets used for compression}
\label{table1}
\resizebox{\textwidth}{!}{
\begin{tabular}{@{}llllllll@{}}
\hline
Dataset      & Original size & RUN        & Species   name       & Technology & Platform                   & SE/PE & Read length \\ \hline
SRR4017489\_1 & 16.5 GB       & SRR4017489 & Homo sapiens         & WCS        & Illumina HiSeq 2000        & PE    & 101,101            \\
SRR6691666\_1 & 316.1 GB      & SRR6691666 & Homo sapiens         & WGS        & TruSeq + Illumina HiSeq X  & PE    & 151,151           \\
ERR753370\_1  & 12.7 GB       & ERR753370  & Homo sapiens         & FINISHING  & Illumina HiScanSQ          & PE    & 101,101            \\
SRR1238539   & 74.2 GB       & SRR1238539 & Homo sapiens         & WGS        & Ion Torrent Proton         & SE    & ~177           \\
SRR6178157   & 37.0 GB       & SRR6178157 & Homo sapiens         & WXS        & Ion Torrent Proton         & SE    & ~135             \\
SRR3479107   & 5.9 GB        & SRR3479107 & Mus musculus         & MBD-Seq    & AB 5500xl Genetic Analyzer & PE+SE & 30,30           \\
SRR3479107\_1 & 21.9 GB       &            &                      &            &                            &       &                     \\
SRR5572323   & 29.8 GB       & SRR5572323 & Mus musculus         & FAIRE-seq  & Illumina HiSeq 2000        & SE    & 76               \\
SRR6240776   & 8.5 GB        & SRR6240776 & Arabidopsis thaliana & ATAC-Seq   & Illumina HiSeq 4000        & PE+SE & ~50,~50          \\
SRR6240776\_1 & 14.9 GB       &            &                      &            &                            &       &                     \\ \hline
\end{tabular}
}

    \begin{tablenotes}
    \footnotesize
    \item \textit{Notes:} \_1 denotes that only one out of the two paired-end FASTQ files was used. RUNs with both SE and PE share some common information. 
    \end{tablenotes}

\end{table*}

\begin{table*}[t]
\caption{Performance of compressors}
\label{table2}
\resizebox{\textwidth}{!}{
\begin{tabular}{lclllllllll}
\hline     
\multirow{2}{*}{Dataset} & \multirow{2}{*}{Orignal size} & \multirow{2}{*}{Length} & \multicolumn{7}{c}{Compression ratio}                                       & \multirow{2}{*}{Gain}    

\\ \cmidrule(r){4-10}
 & \multicolumn{1}{l}{}    
 & \multicolumn{1}{l}{}                     
 & \multicolumn{1}{l}{GTZ} & \multicolumn{1}{l}{LEON} & \multicolumn{1}{l}{HARC} & \multicolumn{1}{l}{PgRC} & \multicolumn{1}{l}{gzip} & \multicolumn{1}{l}{AVS-G} & \multicolumn{1}{l}{AMGC}
                         \\ \hline

SRR4017489\_1             & 16.5GB                        & 101,101                                      & {\color{blue} 81.82} & 23.40 & 30.08 & 77.47                        & 19.30 & 51.64                        & {\color{red} 147.82} & {\color{red} 80.66\%}                      \\
SRR6691666\_1             & 316.1GB                       & 151,151                                      & 73.51                        & 24.70 & 60.79 & {\color{blue} 81.05} & 16.81 & 53.56                        & {\color{red} 98.78}  & {\color{red} 21.88\%}                      \\
ERR753370\_1              & 12.7GB                        & 101,101                                      & {\color{blue} 73.89} & 22.22 & -     & -                            & 15.37 & 53.45                        & {\color{red} 121.20} & {\color{red} 64.03\%}                      \\
SRR1238539                & 74.2GB                        & $\sim$177                                    & {\color{blue} 19.53} & 8.83  & -     & -                            & 11.78 & 10.78                        & {\color{red} 21.82}  & {\color{red} 11.73\%}                      \\
SRR6178157                & 37.0GB                        & $\sim$135                                    & {\color{blue} 17.62} & 11.56 & -     & -                            & 11.21 & 14.29                        & {\color{red} 19.47}  & {\color{red} 10.50\%}                     \\
SRR3479107                & 5.9GB                         & 30,30                                        & 15.28                        & 13.63 & 13.87 & 18.42                        & 7.51  & {\color{blue} 25.64} & {\color{red} 77.96}  & {\color{red} 204.06\%}                     \\
SRR3479107\_1             & 21.9GB                        & 30                                           & 14.60                        & 13.69 & 15.64 & 24.80                        & 8.42  & {\color{blue} 25.26} & {\color{red} 72.81}  & {\color{red} 188.24\%}                     \\
SRR5572323                & 29.8GB                        & 76                                           & {\color{blue} 45.92} & 11.46 & 14.90 & 24.83                        & 9.93  & 34.66                        & {\color{red} 68.14}  & {\color{red} 48.39\%}                      \\
SRR6240776                & 8.5GB                         & $\sim$50                                     & 17.36                        & 16.50 & -     & -                            & 6.07  & {\color{blue} 19.08} & {\color{red} 21.56}  & {\color{red} 13.00\%}                      \\
SRR6240776\_1             & 14.9GB                        & $\sim$50,$\sim$50                            & 29.55                        & 29.07 & -     & -              
              & 15.60 & {\color{blue} 43.67} & {\color{red} 117.82} & {\color{red} 169.80\%}     \\
\hline
\multicolumn{10}{l}{Average gain over the second-best-performing compressors: } & {\color{red} 81.23\%}
 
\\
\hline
\end{tabular}
}

    \begin{tablenotes}
    \footnotesize
    \item \textit{Notes:} For data with different Read Lengths, HARC and PgRC compression failed. The compression ratio is obtained by \textit{total size/compression size}. Given the extremely small percentage of identifiers, the true read compression ratio is about half of the compression ratio given.
    \end{tablenotes}

\end{table*}

The proposed algorithm, AMGC, was evaluated using \href{http://www.avs.org.cn/en/}{AVS-G} as it is an accessible software for us, and then tested on various real datasets. In the following, we will use “AMGC” to refer to both the algorithm and the compressor AVS-G with our algorithm deployed. AMGC’s performance was compared to various existing algorithms. We tested the performance against GTZ, Leon, HARC, PgRC and gzip. GTZ is a very widely used compressor nowadays. Its reference compression mode has a very high compression ratio and compression efficiency. We only test the reference compression mode of GTZ and do not package the used reference part (-r -n). Leon, HARC and PgRC are three representative assembly-based methods. They use different ways of splicing reference genes. Gzip is the classical universal compressor. The tests were run on a machine with an 11th Gen Intel Core i7-11700F 2.50 GHz processor and 128 GB of RAM.

\subsection{Datasets}

The datasets used for evaluation were chosen from the standard test datasets provided by AVS-G, which are listed in Table \ref{table1}. These datasets consist of five Homo sapiens, two Arabidopsis thaliana, and three Mus musculus and include information on the raw sizes, read lengths, sequencing platforms, and sequencing methods used.

Our main targets for comparison are GTZ, AVS-G and AMGC. Since our algorithm focuses only on compressing the read part, we preprocessed the FASTQ files before testing. We simplified the identifier part and set the quality values to the same value, enabling our test results to focus on the read part. As the processing of the different parts in GTZ is integrated, we include the identifiers and quality values parts in the compression output, but this does not affect the comparison between GTZ, AVS-G, and AMGC. It should be noticed that the experiments are conducted in a way which favours the assembly-based algorithms. Because the formatted identifiers and quality values still have a slight impact on GTZ, AVS-G, and AMGC, assembly-based algorithms throw them.

\subsection{Compression size and ratio}

Table \ref{table2} illustrated the compression ratio for different compression tools. We highlighted the best-performing compressors in red and the second-best-performing compressors in blue. It was evident from the table that GTZ and AVS-G had their advantages in different data and were currently the two top compressors. PgRC was the best compressor among assembly-based compressors. It got the second-best performance when compressing data SRR6691666\_1. However, AMGC outperformed all other compressors, achieving an average compression ratio that was 81.23\% higher than the second-best-performing compressor.

\subsection{Time and memory usage}

\begin{table*}[t]
\caption{Time usage of compressors}
\label{table3}
\resizebox{\textwidth}{!}{%
\begin{tabular}{@{}llrrrrrrrrrrrrrr@{}}
\hline
\multirow{2}{*}{Dataset} & \multirow{2}{*}{Orignal size} & \multicolumn{7}{c}{Encoding Time}                                                                                                                              & \multicolumn{7}{c}{Decoding Time}                                                      \\ \cmidrule(r){3-9} \cmidrule(r){10-16}        &               
& \multicolumn{1}{l}{GTZ} & \multicolumn{1}{l}{LEON} & \multicolumn{1}{l}{HARC} & \multicolumn{1}{l}{PgRC} & \multicolumn{1}{l}{gzip}& \multicolumn{1}{l}{AVS-G} & \multicolumn{1}{l}{AMGC}  
& \multicolumn{1}{l}{GTZ} & \multicolumn{1}{l}{LEON} & \multicolumn{1}{l}{HARC} & \multicolumn{1}{l}{PgRC} & \multicolumn{1}{l}{gzip}& \multicolumn{1}{l}{AVS-G} & \multicolumn{1}{l}{AMGC} \\ \hline
SRR4017489\_1            & 16.5 GB                                          & 2m7s                    & 3m47s                    & 5m56s                    & 5m46s                    & 4m8s                     & 3m5s                       & 3m19s                    & 23s                     & 1m20s                    & 41s                      & 17s                      & 39s                      & 1m18s                      & 1m22s                    \\
SRR6691666\_1            & 316.1 GB                                         & 21m18s                  & 126m9s                   & 149m9s                   & 294m5s                   & 65m51s                   & 36m49s                     & 41m33s                   & 21m33s                  & 45m23s                   & 32m6s                    & 13m2s                    & 35m37s                   & 32m52s                     & 32m59s                   \\
ERR753370\_1             & 12.7 GB                                          & 2m34s                   & 3m20s                    & -                        & -                        & 3m42s                    & 1m40s                      & 1m47s                    & 2m5s                    & 1m1s                     & -                        & -                        & 31s                      & 16s                        & 17s                      \\
SRR1238539               & 74.2 GB                                          & 9m31s                   & 41m24s                   & -                        & -                        & 20m10s                   & 4m55s                      & 17m47s                   & 13m6s                   & 12m53s                   & -                        & -                        & 4m18s                    & 5m33s                      & 5m31s                    \\
SRR6178157               & 37.0 GB                                          & 5m39s                   & 11m58s                   & -                        & -                        & 9m42s                    & 4m0s                       & 7m5s                     & 6m30s                   & 4m3s                     & -                        & -                        & 1m40s                    & 1m22s                      & 1m25s                    \\
SRR3479107               & 5.9 GB                                           & 56s                     & 53s                      & 13m7s                    & 3m44s                    & 3m19s                    & 1m55s                      & 1m15s                    & 11s                     & 35s                      & 22s                      & 19s                      & 17s                      & 10s                        & 14s                      \\
SRR3479107\_1            & 21.9 GB                                          & 2m0s                    & 2m47s                    & 78m15s                   & 11m30s                   & 7m57s                    & 1m50s                      & 2m2s                     & 42s                     & 3m20s                    & 42s                      & 57s                      & 1m2s                     & 34s                        & 44s                      \\
SRR5572323               & 29.8 GB                                          & 3m24s                   & 8m33s                    & 16m47s                   & 21m22s                   & 10m45s                   & 3m22s                      & 3m8s                     & 1m4s                    & 3m28s                    & 2m15s                    & 1m19s                    & 1m21s                    & 1m52s                      & 1m54s                    \\
SRR6240776               & 8.5 GB                                           & 1m10s                   & 1m55s                    & -                        & -                        & 5m44s                    & 47s                        & 33s                      & 1m20s                   & 59s                      & -                        & -                        & 26s                      & 14s                        & 16s                      \\
SRR6240776\_1            & 14.9 GB                                          & 1m24s                   & 2m26s                    & -                        & -                        & 3m14s                    & 1m31s                      & 1m10s                    & 1m53s                   & 1m44s                    & -                        & -                        & 36s                      & 16s                        & 19s                      \\
\hline
\end{tabular}
}

    \begin{tablenotes}
    \footnotesize
    \item \textit{Notes:} For data with different Read Lengths, HARC and PgRC compression failed.
    \end{tablenotes}
    
\end{table*}

\begin{table*}[t]
\caption{Memory usage of compressors}
\label{table4}
\resizebox{\textwidth}{!}{%
\begin{tabular}{@{}llrrrrrrrrrrrrrr@{}}
\hline
\multirow{2}{*}{Dataset} & \multirow{2}{*}{Orignal size} & \multicolumn{7}{c}{Encoding RAM}                                                                                                                               & \multicolumn{7}{c}{Decoding RAM}                                           \\ \cmidrule(r){3-9} \cmidrule(r){10-16}        &               
& \multicolumn{1}{l}{GTZ} & \multicolumn{1}{l}{LEON} & \multicolumn{1}{l}{HARC} & \multicolumn{1}{l}{PgRC} & \multicolumn{1}{l}{gzip}& \multicolumn{1}{l}{AVS-G} & \multicolumn{1}{l}{AMGC}  
& \multicolumn{1}{l}{GTZ} & \multicolumn{1}{l}{LEON} & \multicolumn{1}{l}{HARC} & \multicolumn{1}{l}{PgRC} & \multicolumn{1}{l}{gzip}& \multicolumn{1}{l}{AVS-G} & \multicolumn{1}{l}{AMGC} \\ \hline
SRR4017489\_1            & 16.5 GB                                          & 4.0G & 4.8G & 3.3G  & 3.4G  & 1.7M & 9.5G   & 23.4G & 9.0G  & 1.4G  & 0.4G & 2.2G  & 1.4M & 5.4G   & 8.4G  \\
SRR6691666\_1            & 316.1 GB                                         & 6.4G & 7.9G & 47.8G & 72.3G & 1.6M & 13.9G  & 34.7G & 9.6G  & 13.2G & 1.9G & 24.7G & 1.5M & 5.5G   & 23.9G \\
ERR753370\_1             & 12.7 GB                                          & 5.7G & 0.9G & -     & -     & 1.6M & 8.9G   & 21.1G & 17.4G & 1.4G  & -    & -     & 1.4M & 4.0G   & 9.4G  \\
SRR1238539               & 74.2 GB                                          & 9.3G & 6.1G & -     & -     & 1.8M & 11.7G  & 23.0G & 19.4G & 8.0G  & -    & -     & 1.4M & 3.7G   & 11.8G \\
SRR6178157               & 37.0 GB                                          & 8.9G & 2.5G & -     & -     & 1.7M & 8.8G   & 22.6G & 18.6G & 3.5G  & -    & -     & 1.4M & 3.9G   & 12.0G \\
SRR3479107               & 5.9 GB                                           & 3.3G & 0.5G & 0.9G  & 3.3G  & 1.6M & 17.0G  & 18.2G & 7.1G  & 0.6G  & 0.3G & 1.7G  & 1.5M & 3.2G   & 9.4G  \\
SRR3479107\_1            & 21.9 GB                                          & 4.4G & 1.7G & 6.5G  & 11.6G & 1.8M & 9.5G   & 17.4G & 9.4G  & 6.1G  & 0.5G & 4.3G  & 1.5M & 4.4G   & 13.0G \\
SRR5572323               & 29.8 GB                                          & 6.9G & 2.2G & 9.2G  & 10.0G & 1.7M & 12.4G  & 29.9G & 8.6G  & 2.7G  & 0.9G & 4.5G  & 1.4M & 5.2G   & 16.3G \\
SRR6240776               & 8.5 GB                                           & 7.6G & 0.5G & -     & -     & 1.6M & 6.6G   & 12.0G & 12.4G & 1.2G  & -    & -     & 1.3M & 2.7G   & 8.7G  \\
SRR6240776\_1            & 14.9 GB                                          & 6.1G & 0.7G & -     & -     & 1.7M & 3.1G   & 14.4G & 16.1G & 0.4G  & -    & -     & 1.4M & 3.3G   & 10.8G \\
\hline
\end{tabular}
}

    \begin{tablenotes}
    \footnotesize
    \item \textit{Notes:} For data with different Read Lengths, HARC and PgRC compression failed.
    \end{tablenotes}
    
\end{table*}

Tables \ref{table3} and Tables \ref{table4} presented the time and memory usage for compression and decompression. All compressors were run with sixteen threads. For compressors that compressed the whole FASTQ files, the quality values did less impart on compress ratio but still occupied corresponding memory because of the preprocess of FASTQ files.

The reference-based compressors, such as AVS-G and GTZ, showed better time usage when compressing and decompressing. Assembly-based compressors required more time to assemble all input reads into a sequence when compressing, resulting in higher time usage. Compared to AVS-G, AMGC had a similar time complexity, which meant our algorithm did not significantly increase the time required. 

A special example was SRR1238539 from the Ion Torrent Proton platform, where AMGC used more compression time than AVS-G. This was because reads from this platform may be longer than normal NGS reads. The recursive split matching module of AMGC required more alignment times for longer reads, resulting in longer encoding times. However, our algorithm was stable with other data sets. When compared to GTZ, AMGC performs similarly in compression but better in decompression.

In terms of memory consumption, GTZ and Leon were efficient while gzip had a notable advantage. When dealing with large-size data, HARC and PgRC used a lot of memory. AMGC had higher overall memory usage. However, with the development of memory technology, larger memory has become widely available, and the amount of memory used is no longer a significant factor in determining the strengths and weaknesses of genomic compression algorithms, but the compression ratio is.

\subsection{Discussion}

 Even though assembly-based compression algorithms do not require an external reference sequence, the process of read matching exists after assembling their reference sequence. Therefore, our algorithm is suitable for both kinds of methods.
 
Overall, AMGC gets an 81.23\% gain compared with the second-best-performing compressors with comparable RAM and time usage. This is significant work in lossless compression.

\section{Conclusion and future work}

We have introduced AMGC, a new algorithm that includes three modules to boost the compression of reads based on read matching. This algorithm can be integrated into various genomic compression tools. AMGC outperformed state-of-the-art methods and gets an 81.23\% gain compared to the second-best-performing compressor.

Future work could involve extending AMGC to the compression of spatial omics genomic data by developing a more suitable k-mer mapping tool and incorporating spatial omics information into the encoding of mapping results.

\section*{Funding}


This work has been supported by NSFC 61875157 and an open project of BGI-Shenzhen, Shenzhen518000, China.

\bibliographystyle{natbib}
\bibliography{reference}









\end{document}